\documentstyle[pra,aps,twocolumn]{revtex}
\begin{document}
\draft
\author{Li-Bin Fu$^1$ \thanks{%
Email: fu$_-$libin@mail.iapcm.ac.cn }, Jie Liu$^{2,3}$ and
Shi-Gang Chen$^1$}
\address{$^1$Institute of Applied Physics and Computational Mathematics,\\
P.O. Box 8009 (26), 100088 Beijing, China\\
$^2$CAST (World Laboratory), P.O.Box 8730, Beijing, China\\
$^3$ Department of Physics, University of Texas, Austin, Texas 78712}
\title{Critical onset in coherent oscillations between two weakly coupled
Bose-Einstein condensates}
\maketitle

\begin{abstract}
%\begin{center}
%{\bf Abstract }
%\end{center}

%\baselineskip 12pt
The Josephson effects in two weakly linked Bose-Einstein condensates have
been studied recently. In this letter, we study the equations derived by
Giovanazzi et. al. [Phys. Rev. Lett. {\bf 84}, 4521 (2000)] focusing on the
effects of the initial acceleration and the velocity of the barrier on the
``dc'' current. We find that the dc current has lifetime which critically
depends on the moving velocity of the barrier. Moreover, the influence of
the initial acceleration is also investigated and found to be crucial for
the experimental realization of the effects.
\end{abstract}


The Josephson effects (JE's) as a paradigm of the phase coherence
manifestation in a macroscopic quantum system, have been observed in
superconductors \cite{n2}, and demonstrated in two weakly linked superfluid $%
^3$He-{\it B} reservoirs \cite{n4}. Since magnetic and optical traps can be
tailored and biased with high accuracy \cite{n6,n7,n8}, the weakly
interacting Bose-Einstein condensate (BEC) can provide a further context for
JE's and reveal novel properties that might not be accessible with other
systems. Recently the dc and ac Josephson effects in two weakly linked BECs
have been extensively studied \cite{new,n9}. These authors suggested that as
the barrier between the two trapped BECs moves adiabatically across the
trapping potential, a dc current of atoms between two condensates can be
found. In analog of the voltage-current characteristic in superconducting
Josephson junction (SJJ), there exists a critical velocity of the barrier,
at which an abrupt transition from the dc to ac current occurs. In this
letter, we study the model introduced in Ref. \cite{new} focusing on the
effects of the initial acceleration and the velocity of the barrier on the
``dc'' current. We find that the dc current has lifetime which depends on
the velocity of the barrier and is sensitive on the choice of the initial
conditions. To consider the experimental observability of this phenomenon,
we investigate the influence of the initial acceleration and find it plays a
crucial role.

The interacting BECs in a trap at zero temperature can be described by a
macroscopic wave function $\Psi ({\bf {r},}t),$ having the meaning of an
order parameter and satisfying the Gross-Pitaevskii equation
\begin{equation}
i\hbar \frac \partial {\partial t}\Psi ({\bf {r},}t)=[H_0+g|\Psi {\bf ({r},}t%
{\bf )|}^2]\Psi {\bf ({r},}t{\bf ),}  \label{gpe}
\end{equation}
where $H_0=-\frac{\hbar ^2}{2m}\nabla ^2+V_{ext}({\bf {r,}}t{\bf )}$ and $%
g=4\pi \hbar ^2a/m$ with $m$ the atomic mass, and $a$ the s-wave scattering
length of the atoms. Considering the system proposed in Ref. \cite{new}, a
double-well trap produced by a far off-resonance laser barrier, $%
V_{laser}(z)=V_0\exp [-(z-l)^2/\lambda ^2],$ which cuts a single trapped
condensate into two parts \cite{s14}. So, the external potential is given by
the magnetic trap and the laser barrier $V_{ext}({\bf r},t)=V_{trap}({\bf r}%
)+V_{laser}(z,t).$

By solving variationally the GPE using the ansatz: $\Psi ({\bf {r},}t)=\psi
_1(t)\phi _1({\bf r})+\psi _2(t)\phi _2({\bf r})$, where $\psi _{1,2}=\sqrt{%
N_{1,2}(t)}e^{i\theta _{1,2}(t)}$ are complex time-dependent amplitudes, $%
N_{1,2}(t)$ and $\theta _{1,2}$ are the number of atoms and the phase of the
two condensates respectively. The trial wave functions $\phi _{1,2}({\bf r})$
are orthonormal and can be interpreted as approximate ground state solutions
of the GPE of two well respectively. Then the equations of the motion for
the relative population $p(t)=(N_1(t)-N_2(t))/N$ and phase $\theta =\theta
_2-\theta _1$ between the two condensates should be \cite{new,n9}
\begin{equation}
\dot \theta =-Fl+\frac{Kp}{\sqrt{1-p^2}}\cos \theta +Cp,  \label{q1}
\end{equation}
\begin{equation}
\dot p=-K\sqrt{1-p^2}\sin \theta ,  \label{q2}
\end{equation}
where $K=-\frac 2\hbar \int d{\bf r}\phi _1[H_0+gN\phi _1^2]\phi _2{\bf ,}$ $%
C=\frac{gN}\hbar \int |\phi _1|^4d{\bf r}$ and $F=\frac 1\hbar \int d{\bf r}%
[\phi _1^2-\phi _2^2]\frac \partial {\partial l}V_{laser}$ . These equations
describe the dynamics of the Bose Josephson junction (BJJ). For the
convenience we set the parameters as the same as in Ref. \cite{new}, $%
K=4.82\times 10^{-4}$ms$^{-1},$ $C=1.23\;$ms$^{-1}\;$and $F=1.06\;$ms$%
^{-1}\mu $m$^{-1}.$

The Josephson current is defined as the current of atoms across the barrier,
which is $J=\dot N_1=-\dot N_2$. Considering the total number of atoms is
constant , we introduce the normalized current $j=J/N$ . From the Eq. (\ref
{q2}), we have
\begin{equation}
j=\frac{\dot p}2=-\frac K2\sqrt{1-p^2}\sin \theta .  \label{curr}
\end{equation}

Firstly, let us review the properties of this system for $l$ is fixed \cite
{n9}. For this case, \ $p$ and $\theta $ are canonically conjugate variable
of a classical Hamiltonian $H=\frac C2p^2-K\sqrt{1-p^2}\cos \theta -Flp,$
with $\dot p=-\frac{\partial H}{\partial \theta },$ $\dot \theta =\frac{%
\partial H}{\partial p}.$ This system exhibits two qualitatively different
orbits: rotation and libration . In the rotation regime, $\theta $ increases
(or decreases) monotonically and $p$ oscillates with small amplitude. In the
libration regime, $\theta $ and $p$ oscillates around the equilibrium point $%
P_e(p_e,\theta _e)$. The separatrix between the two regimes determined by
the saddle point $P_s(p_s,\theta _s)$ with $H_s=H(p_s,\theta _s).$ For large
energy $(H>H_s)$ the orbit is rotation, whereas for small energy $(H<H_s)$
the orbit is libration. The equilibrium point and the saddle point are
obtained by equating the right hand sides of Eqs. (\ref{q1}) and (\ref{q2})
to zero, yielding $p_e\approx \frac{Fl}C,\theta _e=0$ and $p_s\approx \frac{%
Fl}C,\theta _s=\pi .$

In the rotation regime, since $\theta $ is increasing (or decreasing)
monotonically with the angular frequency proportion to $\sqrt{KC}$ , from
equation (\ref{curr}), we find that $j$ is a fast oscillation current with
frequency proportion to $\sqrt{KC}$ and the upper bound of the current,
i.e., the so-called critical current $j_c\approx \frac K2\sqrt{1-p_a^2}$
where $p_a$ is the average of $p.$

In the liberation regime, let $p=p_e+\delta p,$ considering $C\gg K$ , from
Eqs. (\ref{q1}) and (\ref{q2}) we can obtain a pendulum equation: $\ddot
\theta +KC\sqrt{1-p_e^2}\sin (\theta )=0$. Then, we obtain $\theta =\theta
_p\sin (\omega t),\;\;\delta p=\frac{\omega \theta _p}C\cos (\omega t)$ in
which $\omega =\sqrt{KC(1-p_e^2)}$ and $\theta _p$ is determined by the
energy of the pendulum. So, one gets $\delta p=\theta _p\sqrt{\frac KC%
(1-p_e^2)}\sim $ $\sqrt{\frac KC}\approx 10^{-2}.$ For this case, the
Josephson current is $j=-\frac K2\sqrt{1-p_e^2}\sin [\theta _p\sin (\omega
t)].$ The critical current is $j_c=\frac K2\sqrt{1-p_e^2}\sin (|\theta _p|)$
or $j_c=\frac K2\sqrt{1-p_e^2}$ when $|\theta _p|\geq \pi /2$.

From the above discussion, for a fixed $l$ one can only find the ac current
in BJJ. These properties exhibit the analog of the dc voltage case in SJJ.

As it has been suggested in Ref. \cite{new}, a dc current can be induced by
moving the laser barrier across the trap with a constant velocity $V=\frac{dl%
}{dt},$ and exhibit the analog of the critical behavior in SJJ's. Now the
question arises: can this system exhibit some new properties which can not
find in SJJ? To answer this question, let us review the case for SJJ. In
SJJ, we know that the densities of cooper pair in two side of the junctions
are equal to each other (if the materials are the same) and should hardly
change when the dc current exists. This feature interpreting to the BJJ's is
that the change in the relative population is very small. To see this
feature clearly, let us assume the initial relative population $p(0)=0$,
since the change in the relative population is very small, the Eqs. (\ref{q1}%
) and (\ref{q2}) can be approximated to
\begin{equation}
\dot p=-K\sin (\theta )  \label{qqq1}
\end{equation}
\begin{equation}
\;\dot \theta =-FVt+Cp.  \label{qqq2}
\end{equation}
Differentiating the second of these equations and replacing the first, we
have $\ddot \theta +KC\sin \theta =-FV,$ which is a driven pendulum
equation. The first integral of the equation gives the energy of this
pendulum $E=\frac 12\dot \theta ^2-KC\cos \theta +FV\theta ,$ and the
constant $E=\frac 12\dot \theta (0)^2-KC\cos \theta [(0)]+FV\theta (0).$ We
define the potential
\begin{equation}
U(\theta )=-KC\cos \theta +FV\theta .  \label{poten}
\end{equation}
When $V<0.559\mu m/s,$ i.e., $FV/(KC)<1,$ it is a washboard potential, its
local maximum appears at $\theta =(2m-1)\pi +\arcsin k$, and the local
minimum appears at $\theta =2m\pi -\arcsin k$, where $k=\frac{FV}{KC}$ and $%
m $ is an integer$.$ We restrict the initial $\theta $ in interval $[-\pi
,\pi ],$ then the motion of $\theta $ is characterized by the local maximum $%
U_c=KC\sqrt{1-k^2}+FV[\arcsin k-\pi ].$ If $E<U_c,$ the motion of $\theta $
is an oscillation in the interval $[\theta _{\max },\theta _{\min }]$ with
the frequency $\omega \approx \sqrt{KC}$, where $\theta _{\max }$ and $%
\theta _{\min }$ is the solution of $E=U(\theta )$. If $E>U_c,$ the motion
of $\theta $ is a rotation, i,e., $\theta $ decreases monotonically, and $%
\dot \theta =\sqrt{2(E+KC\cos \theta -FV\theta )}$. When $V>0.559\mu m/s,$
i.e., $FV/(KC)>1,$ the potential is a titled-step potential, $\theta $ is
also a rotation with $\dot \theta \sim -FVt.$ From (\ref{curr}), we know it
is an ac current when $\theta $ is a rotation.

To investigate the case when $\theta $ is in oscillation regime, let $%
p=p_e(t)+\delta p$ where $p_e(t)=\frac{FV}Ct,$ from (\ref{qqq2}), we obtain$%
\;\delta p\approx \frac{\dot \theta }C$, so, we know that $\delta p$ is a
small term with $\delta p\sim \sqrt{\frac KC}$ $\approx 0.02.$ Then from (%
\ref{curr}) we can obtain a dc current (to zero order of $\delta p$)
\begin{equation}
j=\frac{FV}{2C}.
\end{equation}

From the above discussion, when $\theta $ is in oscillation regime, the
current is a dc current with the relative population increasing $(p\approx
p_e(t));$ when $\theta $ is in rotation regime, the current is an ac one.
The motion of $\theta $ is determined by the initial conditions and the
driven force $\frac{FV}C.$ If sets the initial conditions: $\theta (0)=\dot
\theta (0)=0,$ one can obtain a critical velocity $V_c=0.406\mu m/s.$ For $%
V<V_c,$ then $E<U_c$ , it is a dc current, but for $V>V_c,$then $E>U_c,$ the
current is in ac regime. This is a close analog of the critical behavior in
SJJ.

However, for the BJJ's, the dc current must lead to the change of the
relative population. This feature will give rise to new properties. Taking
the change of the relative population into account, we obtain a nonrigid
driven pendulum equation
\begin{equation}
\ddot \theta +KC\sqrt{1-p^2}\sin \theta =-FV.  \label{epe}
\end{equation}
In analog to the above discussion, we get the potential
\begin{equation}
U(p,\theta )=-KC\sqrt{1-p^2}\cos \theta +FV\theta .  \label{poten1}
\end{equation}
This is also a tilted washboard potential for $V<0.559\mu m/s$. The new
feature is the local maximum $U_c(p)$ will decrease with $|p|$ increasing$.$
If the motion of $\theta $ is an oscillation at the beginning $(U_c(p)>E)$,
we will firstly find a dc current with $U_c(p)$ decreasing. The oscillation
of $\theta $ will not keep when $U_c(p)\leq E,$ at this moment $\theta $
will start rotating, and then the current will transition to an ac one. This
means the dc current does not keep for all the time, i.e., the dc current
has a lifetime. So one can have a straightforward definition of lifetime of
a dc current as: the lifetime of the dc current $\tau _c$ is the time at
which $E\geq U_c(p(\tau _c)).$ This definition is consistent with the
definition of the critical velocity in SJJ's.

The voltage-current characteristic is the most important property in SJJ's.
But in BJJ, the important physical quantity should be the relative
population $p$ which can be directly detected. So, the critical behavior
should be characterized by the change of $p$ in BJJ. When $\theta $ is in
oscillation regime, from the above discussion the relative population is
increasing. When $\theta $ is rotating, if $\theta <(\arcsin k-\pi ),$ $\dot
\theta \approx \sqrt{2(E+KC\cos \theta -FV\theta )}$ , $\delta p$ is still a
small term with $\delta p_{\max }\sim \sqrt{\frac KC}$ $\approx 0.02,$ hence
the relative population is still increasing and $p\approx p_e(t)$ (in the
zero order of $\delta p$)$.$ But if $\theta >(\arcsin k-\pi ),$ $\dot \theta
$ will decrease monotonically, $\dot \theta \approx -FVt,$ then $\theta
\approx -\frac 12FVt^2$, the integral of the current: $\int_{t_0}^\infty
K\sin (\frac 12FVt^2)dt\approx 0.0175$ where $t_0$ is a finite time, so $p$
should hardly increase. This means that when the motion of $\theta $ changes
from oscillation to rotation, the relative population will still increase
until $\theta \geq (\arcsin k-\pi ).$ So, the more accurate definition of
the lifetime of the dc current in BJJ should be: the lifetime $\tau _p$ is
the time when $\theta (\tau _p)\geq (\arcsin k-\pi )$, within this time the
relative population is increasing, but after this time the relative
population will hardly increase and keep on average fixed$.$

In Fig. 1, we show the phase diagram relating to the critical onset in the
parameter space of $t$ and the laser velocity $V$ for the initial
conditions: $l(0)=0,\;$ $p(0)=0\ $and$\;\theta (0)=0.$ The solid line is the
lifetime $\tau _p(V),$ the dashed line is the lifetime $\tau _c(V)$. The
step structure implies that the transitions occur in different cycle of the
oscillation. The abruptly increase is due to that the transition occurs near
the peak of the potential where $\dot \theta \approx 0.$ Fig. 2. plots the
relative population after the dc current is destroyed, the crosses are the
relative population $p$ at $t=2s$ which are obtained by integrating the Eqs (%
\ref{q1}) and (\ref{q2})$,$ the solid line is the theoretical result given
by $p=p_e(\tau _p),$ which shows the theoretical estimation is consistent
with the numerical simulation.

From the above discussion, we know that the lifetime of a dc current is also
determined by the initial conditions$.$ One knows that the initial relative
population $p(0)$ must be very close to the equilibrium $p_e$ to obtain a dc
current, but the initial relative phase can be various. In Fig. 3, we plot
the lifetime of the dc current for different initial relative phase $\theta
(0)$, where we let the initial population $p(0)=0$ and $l(0)=0$.

Concerning a possible realization of the phenomenon described in this work,
one should consider the influence on the initial accelerations. We choose
the initial conditions as: $l(0)=0,$ $p(0)=0,\theta (0)=0$ and $V(0)=0.$ Let
the barrier starts moving with a constant initial acceleration. When its
velocity reaches the value $V_0,$ we stop accelerating and then keep the
velocity. What we concern is the change on the lifetime of dc current caused
by the different initial acceleration. Fig. 4(a) shows the lifetime for
different initial acceleration, where the lifetime is the time duration of a
dc current after the acceleration. The solid line is for $V_0=0.5$ $\mu m/s$
and the dashed line is for $V_0=0.4$ $\mu m/s.$ We find that for a sudden
change of the initial barrier, the influence of the initial accelerations
can be negligible, whereas for slow acceleration process the lifetime
changes abruptly. The reason can be given by the following analysis: For a
sudden acceleration, the time for accelerating is very short. In this time
period , the change of $\theta $ is small, so does the change of lifetime.
On the contrary, in a slowly acceleration process, the $\theta $ changes
greatly (see Fig. 4(b)). Because the lifetime is very sensitive to the
initial conditions, so the lifetime will change dramatically.

We note that the conserved conditions implies that $l$ must be less than $%
\frac CF\approx 1.16\mu m$ to ensure $p_e<1$. In another aspect, to ensure
the number of the condensed atoms in one well beyond the minimum threshold, $%
l$ has to be less than this value. On the other hand, although the Eqs. (\ref
{q1}) and (\ref{q2}) is obtained by solving variationally the GPE (\ref{gpe}%
) under the approximation $p$ $<<1,$ it still remains a good approximation
even for $p\approx 0.4$ \cite{new}. So, the discussion in this paper is
accurate at least under these constraints.

We thank Prof. B.Y. Ou for useful discussions. This project was supported by
Fundamental Research Project of China.

\section*{Figures caption}

Fig. 1. The phase diagram relating to the critical onset in the parameter
space of $t$ and the laser velocity $V$ for the initial conditions: $%
l(0)=0,\;$ $p(0)=0\ $and$\;\theta (0)=0.$ The solid line is the lifetime $%
\tau _p(V),$ the dashed line is $\tau _c(V)$.

Fig. 2. The relative population after the dc current is destroyed.

Fig. 3. The lifetime of the dc current for different initial $\theta .$ The
initial relative population is $p(0)=0$ and $l(0)=0.$

Fig. 4(a). The lifetime for different initial accelerations. The solid line
is for $V_0=0.5$ $\mu m/s$ and the dashed line is for $V_0=0.4$ $\mu m/s$.
(b). The relative phase $\theta $ when the velocity reaches $V_0.$

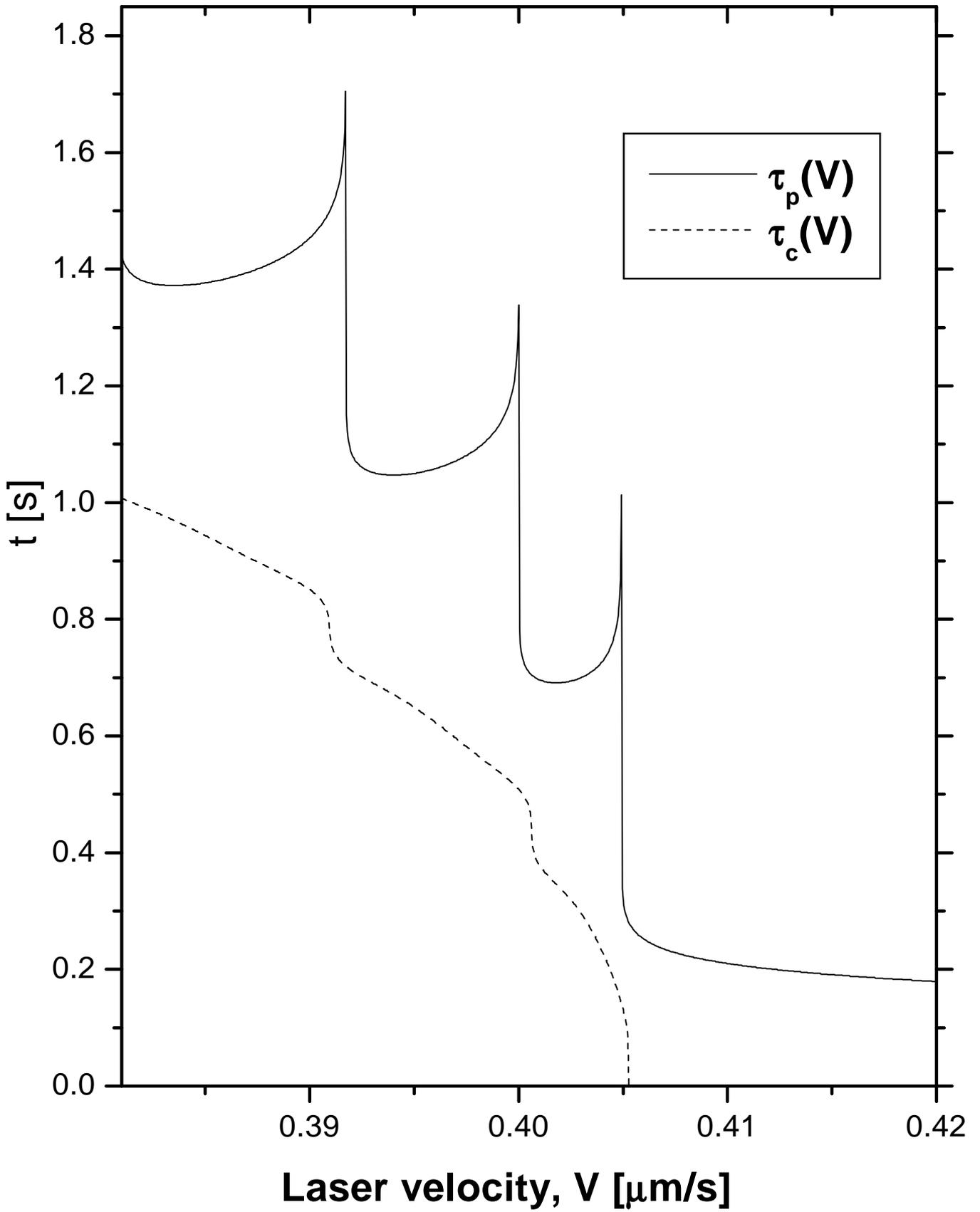
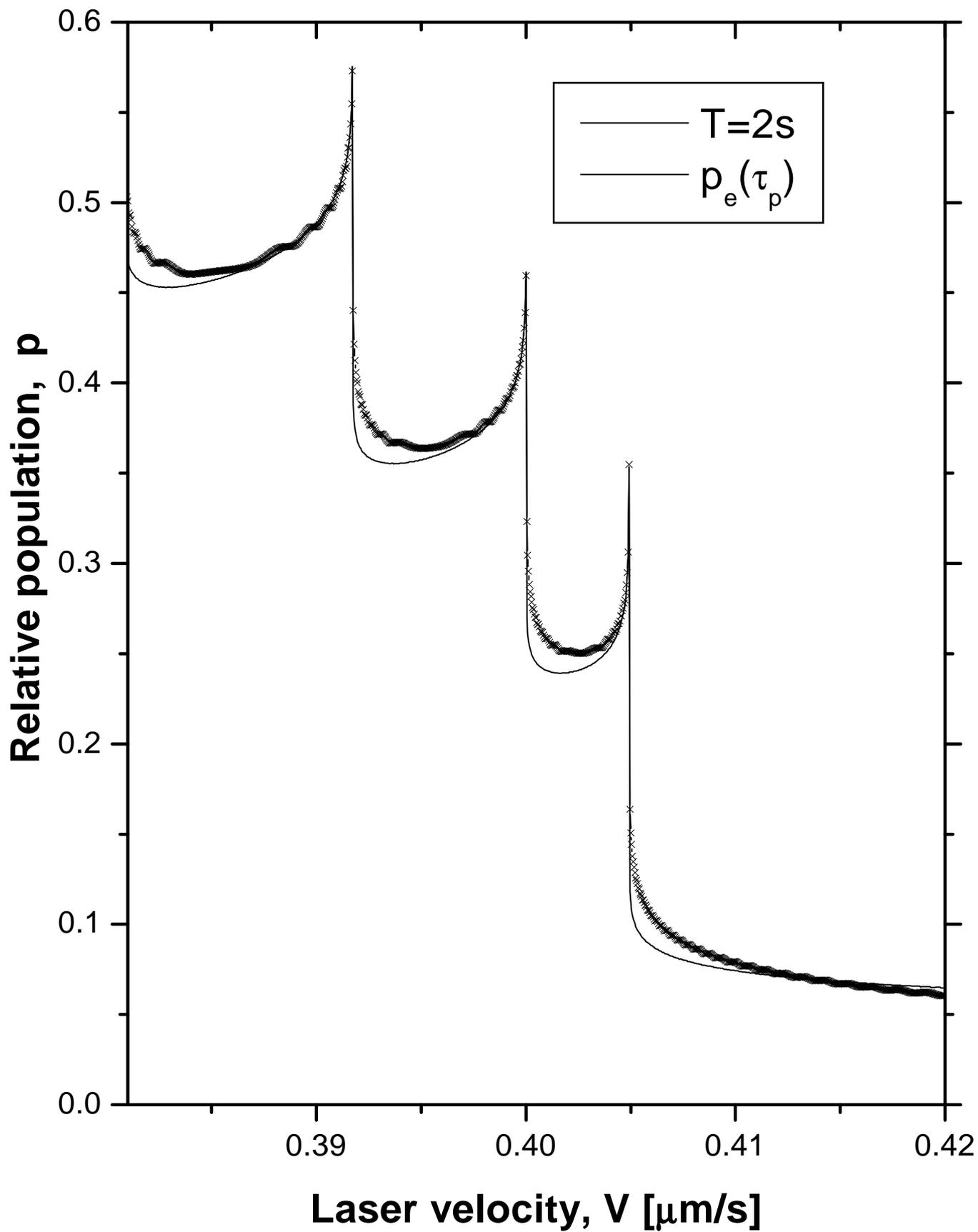
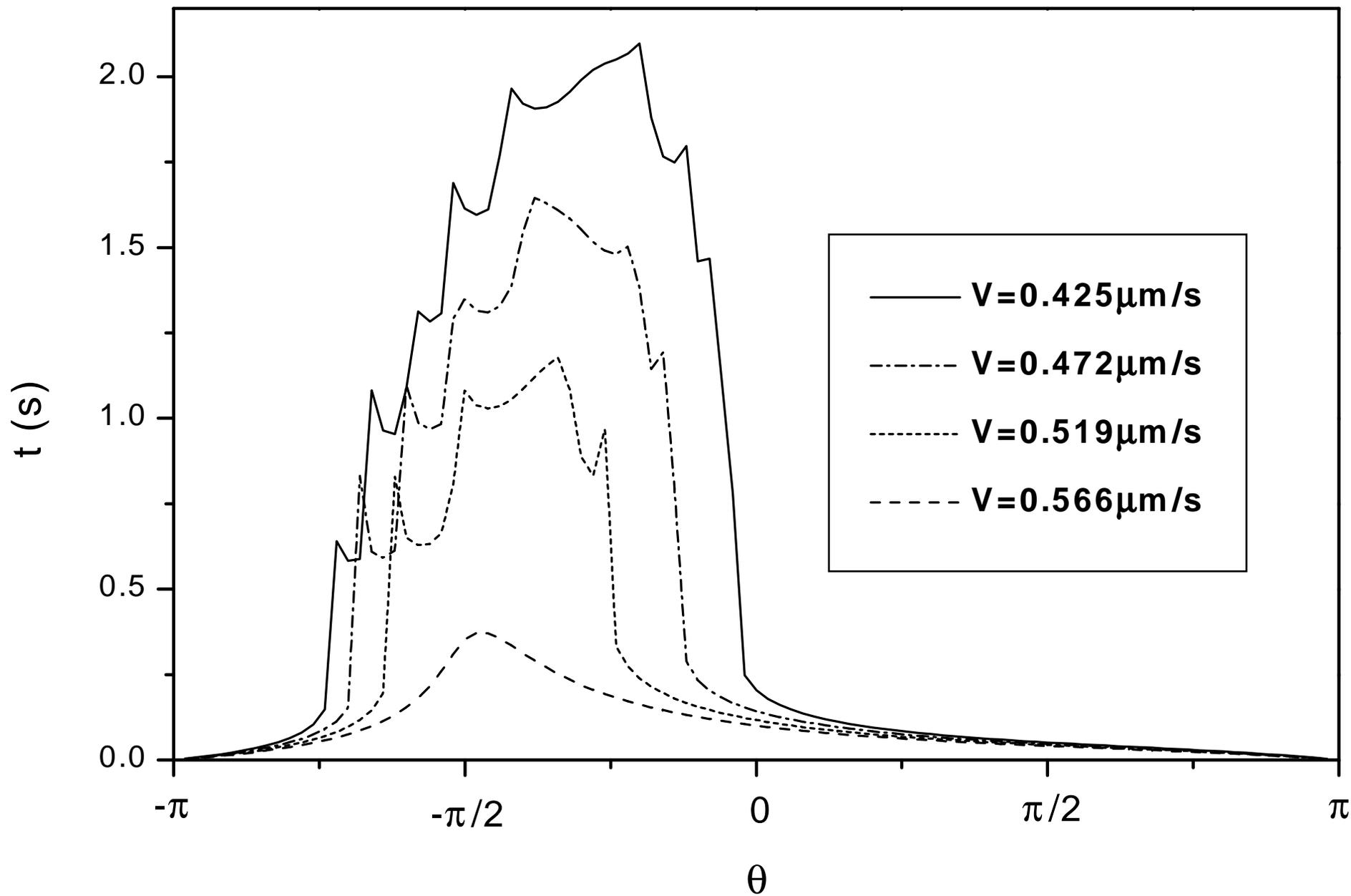
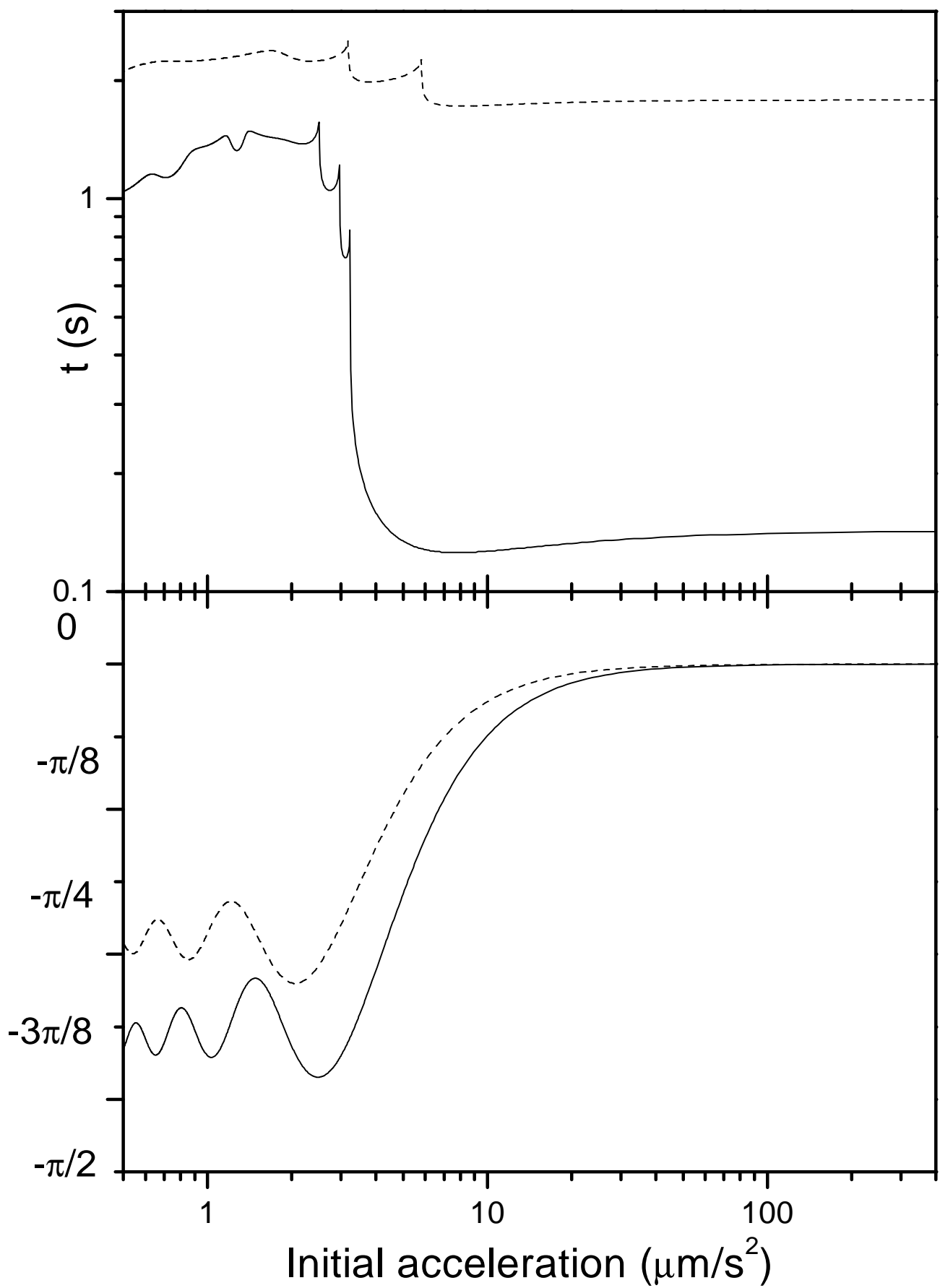
\end{document}